\documentstyle[psfig]{l-aa}
\chardef\PBii="10
\def\PBi{\'\PBii}
\def\PBvp #1 #2{ #1, #2}
\def\PBa #1:#2 #3 #4 {#1, #2, {A\&A} \PBvp #3 #4}
\def\PBna #1:#2 #3 #4 {#1, #2, {NewA} \PBvp #3 #4}
\def\PBapj #1:#2 #3 #4 {#1, #2, {ApJ} \PBvp #3 #4}
\def\PBasupl #1:#2 #3 #4 {#1, #2, {A\&AS} \PBvp #3 #4}
\def\PBapjsupl #1:#2 #3 #4 {#1, #2, {ApJS} \PBvp #3 #4}
\def\PBpasp #1:#2 #3 #4 {#1, #2, { PASP} \PBvp #3 #4}
\def\PBpaspc #1:#2 #3 #4 {#1, #2, { PASPC } \PBvp #3 #4}
\def\PBmn #1:#2 #3 #4 {#1, #2, {MNRAS} \PBvp #3 #4}
\def\PBobs #1:#2 #3 #4 {#1, #2, {The Observatory} \PBvp #3 #4}
\def\PBprsoc #1:#2 #3 #4 {#1, #2, {Proc. R. Soc. Lond. A.} \PBvp #3 #4}
\def\PBmsait #1:#2 #3 #4 {#1, #2, {Mem. S.A.It.} \PBvp #3 #4}
\def\PBnat #1:#2 #3 #4 {#1, #2, {Nat} \PBvp #3 #4}
\def\PBaj #1:#2 #3 #4 {#1, #2, {AJ} \PBvp #3 #4}
\def\PBjaa #1:#2 #3 #4 {#1, #2, {JA\& A} \PBvp #3 #4}
\def\PBaspsc #1:#2 #3 #4 {#1, #2, {Ap\&SS} \PBvp #3 #4}
\def\PBanrev #1:#2 #3 #4 {#1, #2, {ARA\&A} \PBvp #3 #4}
\def\PBrevmex #1:#2 #3 #4 {#1, #2, {Rev. Mex. de Astron. y Astrof.} \PBvp #3 #4}

\def\PBscie #1:#2 #3 #4 {#1, #2, {Sci} \PBvp #3 #4}
\def\PBesomsg #1:#2 #3 #4 {#1, #2, {The Messenger} \PBvp #3 #4}
\def\PBrmp #1:#2 #3 #4 {#1, #2, {Rev. Mod. Phys.} \PBvp #3 #4}
\def\PBans #1:#2 #3 #4 {#1, #2, {Ann. Rev. of Nucl. Sci.} \PBvp #3 #4}
\def\PBphrev #1:#2 #3 #4 {#1, #2, {Phys. Rev.} \PBvp #3 #4}
\def\PBphreva #1:#2 #3 #4 {#1, #2, {Phys. Rev. A} \PBvp #3 #4}
\def\PBphs #1:#2 #3 #4 {#1, #2, {Physica Scripta} \PBvp #3 #4}
\def\PBjqsrt #1:#2 #3 #4 {#1, #2, {J. Quant. Spectrosc. Radiat. Transfer} \PBvp 
#3 #4}
\def\PBcjp #1:#2 #3 #4 {#1, #2, {Can. J. Phys. } \PBvp #3 #4}
\def\PBjphb #1:#2 #3 #4 {#1, #2, {J. Phys. B} \PBvp #3 #4}
\def\PBapop #1:#2 #3 #4 {#1, #2, {Appl. Opt.} \PBvp #3 #4}
\def\PBgca #1:#2 #3 #4 {#1, #2, {Geochim. Cosmochim. Acta}\PBvp #3 #4}
\def\PBjmols  #1:#2 #3 #4 {#1, #2, {J. Mol. Spec.}\PBvp #3 #4}
\def\PBchphyl  #1:#2 #3 #4 {#1, #2, {Chem. Phys. Lett.}\PBvp #3 #4}
\def\PBkms{$\rm km s^{-1}$}
\def\PBt{${T}_{\rm eff}~$}
\def\PBg{$\rm \log g$}
\def\PBal{{ et al.~}}
\begin{document}
\thesaurus{08(01.1,16.3,12.1), 10(08.1)}
\title{CS 22957-027: A carbon-rich extremely-metal-poor star
\thanks{This paper is based on observations collected at ESO, La Silla}}
\author{
P. Bonifacio\inst{1}
\and P. Molaro\inst{1}
\and T.C. Beers\inst{2}
\and G. Vladilo\inst{1}}
\offprints{P. Bonifacio}
\institute{Osservatorio Astronomico di Trieste Via G.B. Tiepolo 11 34131 
Trieste, Italy
\and Department of Physics and Astronomy, Michigan State University, East
Lansing MI 48824, USA}
\date{Received July 15th 1997; accepted December 17th 1997}
\maketitle
\begin{abstract}
 
We present a high-resolution spectroscopic analysis of the extremely-metal-poor
star CS 22957-027, which  Beers \PBal (1992) noted to exhibit a rather
strong G-band.  For \PBt = 4839 K, derived from broadband photometry, our
analysis obtains \PBg = 2.25, and a very low metallicity of [Fe/H]=$-3.43\pm
0.12 $.  The carbon-to-iron ratio is found to be enhanced by $\approx 2$ dex
relative to the solar value, similar to the handful of other carbon--rich
metal--deficient stars discussed by Norris \PBal (1997a) and Barbuy \PBal
(1997).  From the $\rm ^{12}C^{13}C$ (1,0) Swan bandhead and $\rm ^{13}CH$
lines the isotope ratio $\rm ^{12}C/^{13}C$ is $\approx$ 10, indicating a $\rm
^{13}C$ enrichment.  Nitrogen is also found to be enhanced by $\approx 1$ dex.
The $s-$ process elements Sr and Ba are, surprisingly, found to be
under-abundant relative to solar, with [Sr/Fe]=-0.91 and [Ba/Fe]=-0.93.  The
star's low luminosity requires that the chemical enrichment arises from mass
transfer from an evolved companion, rather than self-polluting dredge-up
processes.  However, the $s-$ process elements Sr and Ba are unusually low,
with a solar ratio to one another, at variance with what found in the
``classical'' CH stars.
 
Finally we note that a feature due to $^{13}$CH is present in coincidence with
the Th II 401.9129 nm resonance line.  The blend is of relevance for the
nucleo-chronology use of the Th II line in these extremely metal poor stars,
when significant $\rm ^{13}C$ is present.
  
\keywords{Stars:abundances - Stars:Population II - Stars:late type -
Galaxy:halo}
\end{abstract}
 
\section{Introduction}

The surveys of Beers \PBal (1985, 1992, hereafter BPS), aimed at the discovery
of very metal--poor stars, have shown that about 10\% of the stars more
metal--poor than [Fe/H]=--2.0 have a stronger than normal G-band.  At higher
metallicities, the strong--G--band stars are less frequent, on the order of a
few percent (Norris \PBal 1997a).  The excess of carbon-enhanced stars at
progressively lower metal abundance is  unexplained, but is probably linked to
the early stages of Galactic chemical evolution.
 
\begin{figure*}[t]
\psfig{figure=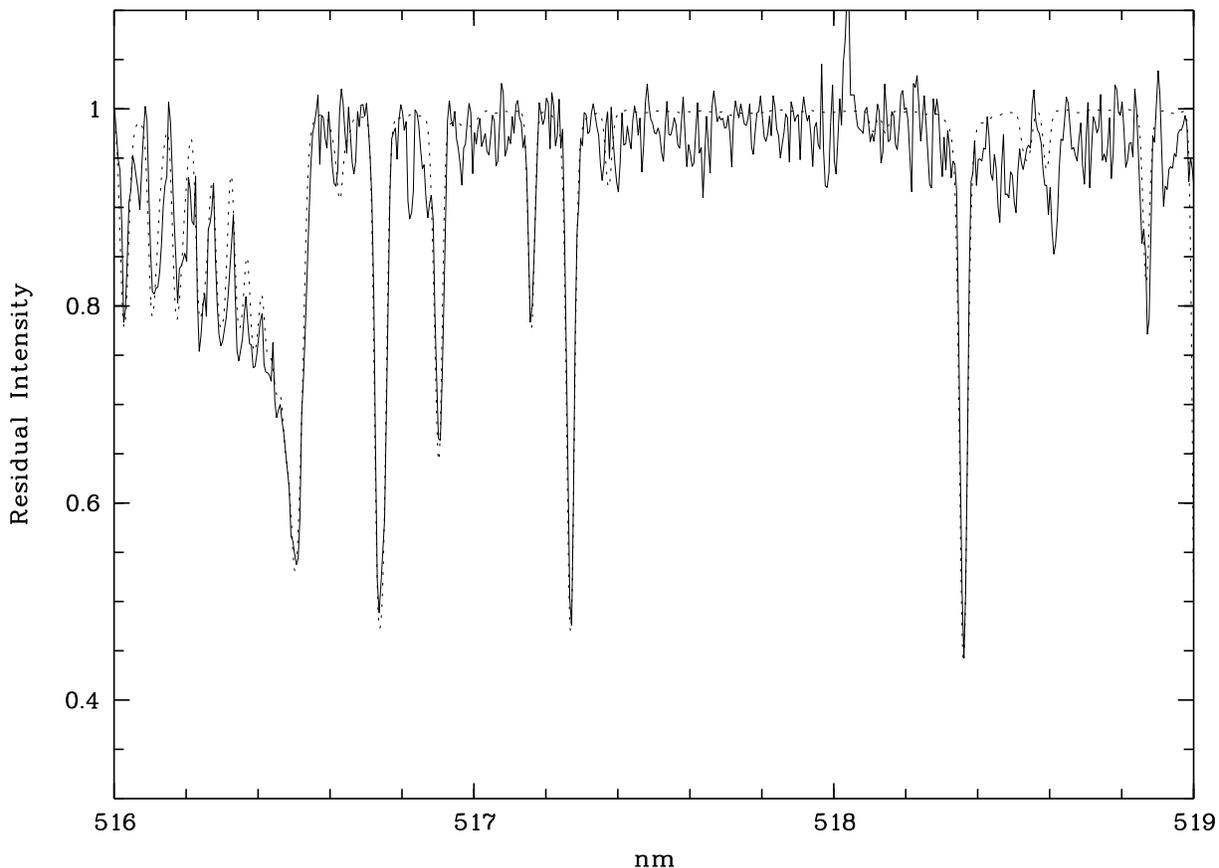,width=18cm,angle=-90}
\caption[]{The 
C$_2$ (0,0) Swan bandhead and the Mg I b triplet.  The dotted line is a
synthetic spectrum with [C/Fe]=+2.25 .  }
\end{figure*}
 
\par
{
\renewcommand{\tabcolsep}{3pt}
\begin{table}[t]
\caption{Log of Observations}
\begin{center}
\begin{tabular}{cccccrl}
Date & Tel. & Texp & $\lambda$ range & S/N & R & $v_r$\\
     &      & sec    & nm              &   &   & \PBkms  \\
\hline
1981/06/08&PAL5.0 & 180 & 380-450 & 20 & 4000 & $-61\pm 15$\\
1993/09/23&ESO1.5 & 600 & 380-450 & 20 & 4000 & $ -78\pm 10$\\
1996/09/05&ESO1.5 & 900 & 380-450 & 25 & 4000 & $-74\pm 10 $\\
1996/09/27&ESO3.6 & 1810 & 393-523 & 20 & 32000 & $-72\pm 2$\\
1996/09/28&ESO3.6 & 3630 & 350-463 & 30 & 32000 & $-74\pm 2$\\
\hline
\end{tabular}
\end{center}
\end{table}
}
{\renewcommand{\arraystretch}{0.7}
\begin{table}[b]
\caption{Atomic line data and abundances ($\epsilon = \log {\rm (X/H)} + 12$) }
\begin{center}
\begin{tabular}{llrrrc}
\hline 
\\
$\lambda$ & ion &log gf & EW &$\epsilon$ & Comment\\
 (nm)     &     &       & (pm)&  &   \\
\hline
\\
457.1096 & Mg I & -5.69 & 1.15 &  4.35 & syn \\
470.2991 & Mg I & -0.67 & --   &  4.88 & syn \\
516.7321 & Mg I & -1.03 & --   &  4.48 & syn \\
517.2684 & Mg I & -0.42 & --   &  4.48 & syn \\
518.3604 & Mg I & -0.18 & --   &  4.48 & syn \\
422.6728 & Ca I &  0.24  & --   &  3.36 & syn \\
443.4957 & Ca I & -0.03 & --   &  3.36 & syn \\
393.3663 & Ca II&  0.13 & --   &  3.36 & syn \\
396.8469 & Ca II& -0.17 & --   &  3.36 & syn \\
390.4785 & Ti I &  0.03 & 3.80 &  2.82 & \\
392.4527 & Ti I & -0.94 & 1.76 &  2.28 & \\
400.8926 & Ti I & -1.04 & 2.20 &  2.49 & \\
468.1908 & Ti I & -1.00 & 1.10 &  2.05 & \\
498.1732 & Ti I &  0.52 & 3.18 &  2.03 & \\
499.1067 & Ti I &  0.35 & 1.21 &  1.65 & \\
502.2871 & Ti I & -0.39 & 1.23 &  2.39 & \\
503.9959 & Ti I & -1.19 & 1.00 &  2.14 & \\
374.1635 & Ti II& -0.14 & 8.00 &  2.33 & \\
398.7600 & Ti II& -2.63 & 2.19 &  2.31 & \\
416.3648 & Ti II& -0.40 & 1.88 &  2.30 & \\
444.3794 & Ti II& -0.81 & 8.11 &  2.21 & \\
450.1273 & Ti II& -0.75 & 6.51 &  1.83 & \\
456.3761 & Ti II& -0.95 & 6.64 &  2.17 & \\
518.8680 & Ti II& -1.20 & 5.58 &  2.57 & \\
359.3481 & Cr I & 0.31  & --   &  2.27 & syn \\
390.8756 & Cr I & -1.00 & --   &  2.77 & syn\\
425.4332 & Cr I & -0.11 & --   &  2.07 & syn\\
464.6148 & Cr I & -0.70 & 1.41 &  2.24 & \\
520.4506 & Cr I & -0.17 & --   &  1.87 & syn \\
520.6038 & Cr I & 0.03 & --    &  1.97 & syn \\
520.8419 & Cr I & 0.16 & --    &  1.97 & syn \\
362.3183 & Fe I & -0.71 & 3.05 &  3.92 & \\
368.5998 & Fe I & -0.23 & 4.56  & 4.35 & \\
371.9935 & Fe I & -0.43 & 21.40 & 3.99 & \\
382.1179 & Fe I & 0.24 & 4.66   & 4.26 & \\
390.6479 & Fe I & -2.24 & 8.94 &  4.13 & \\
422.2213 & Fe I & -0.93 & 2.67 &  4.01 & \\
423.3602 & Fe I & -0.56 & 4.51 &  4.05 & \\
426.0473 & Fe I & 0.13  & 8.11 &  4.05 & \\
492.0502 & Fe I &  0.07 & --   &  3.96 & syn  \\
495.7298 & Fe I & -0.34 & --   & 3.91  & syn \\
495.7597 & Fe I & 0.13  & --   &  4.01 & syn \\
501.2067 & Fe I & -2.64 & --   & 4.06 & syn \\
504.1755 & Fe I & -2.09 & --   & 3.96 & syn \\
517.1595 & Fe I & -1.79 & --   & 4.01 & syn \\
519.1455 & Fe I & -0.66 & --   & 4.16 & syn \\
520.2335 & Fe I & -1.56 & --   & 3.91 & syn \\
523.2939 & Fe I & -0.14 & 4.40 & 4.05 & \\
458.3837 & Fe II & -2.02 & 3.35 & 4.40 & \\
492.3927 & Fe II & -1.32 & 6.24 & 4.33 & \\
366.4085 & Ni I & -2.07 & 5.12 &  2.89 & \\
367.0423 & Ni I & -2.08 & 5.00 &  2.74 & \\
377.2525 & Ni I & -3.06 & 3.03 &  3.35 & \\
407.7709 & Sr II&  0.17 & --  &   -1.42& syn \\
421.5519 & Sr II& -0.15 & --  &   -1.47& syn  \\
455.4029 & Ba II& 0.17  &  --  &  -2.10& syn \\
493.4076 & Ba II& -0.15&  --  &   -2.37& syn \\
\hline 
\end{tabular}
\end{center}
\end{table}
}
 
Chemical peculiarities in cool stars ($ (B-V)>0.4$) are often interpreted as a
result of mixing nucleosynthesis products to the stellar surface. The
nucleosynthesis may have taken place either in the star itself or in an evolved
companion from which mass has been accreted either through Roche-Lobe overflow
or through stellar winds.  The evolution in a binary system is probably
responsible for the formation of Ba stars, CH-stars, sgCH-stars (McClure, 1984;
McClure \& Woodsworth, 1990; McClure 1997b) and dwarf Carbon stars (Green \&
Margon 1994).  The BPS strong-G-band stars have been labeled ``CH--stars'' in
the BPS papers, but their relation to the more metal--rich CH stars
([Fe/H]$\approx -1.5$), remains to be verified.  Few of these have, to date,
been subject to detailed analysis: CS 22892-052 (McWilliam \PBal 1995, Sneden
\PBal 1996), CS 22948-027, CS 29497-034 (Barbuy \PBal 1997), LP 706-7 and LP
625-44 (Norris\PBal  1997a).  All these stars proved to be enhanced in
$s$--process elements, like the  ``classical'' CH and Ba stars.  
In addition CS
22892-052 shows a unique signature:  the enhancement of $r$--process elements
(e.g.  [Eu/Fe]=+1.7), showing that their production took place early in the
Galaxy.  Sneden \PBal (1996) and McWilliam \PBal (1995) suggested that the
$r-$process enhancement observed in CS 22892-052 originated from the ejecta of
an early supernova.  The $r-$ process element distribution is important to
infer the number and type of events which originated it (see the discussion by
Goriely \& Arnould 1997).  However, the supernova hypothesis leaves the carbon
excess ([C/Fe]=+1) unexplained.  Norris \PBal (1997a) speculated on the
existence of a 10 M$\odot$ star which has gone through an AGB evolution and
ends as a Type II SN, to synthesize both C and $r-$process elements.  The same
authors mention the more exotic possibility of ``hypernovae'' of mass in
excess of 100 M$\odot$.  The presence of Th and other $r-$process elements in
CS 22892-052 has been used to derive a radioactive-decay age of the universe
$t=15.2\pm3.7$ (Cowan \PBal 1997).  Such large enhancements of the $r-$process
elements are certainly not present in LP 706-7 and LP 625-44, but remain to be
investigated for CS 22948-027 and CS 29497-034.  Further studies of this class
of stars should eventually lead to a better understanding of their nature.  In
this paper we describe the analysis of another C-enhanced metal-poor star: CS
22957-027.
 
\section{Observations and data reduction}
 
In September 1996 we collected two  spectra of CS 22957-027 with the CASPEC
spectrograph at the ESO 3.6 m telescope.  The echelle grating with 31.6
lines/mm and the long camera (f/3) were used.  The detector was the Tek CCD
$1024\times 1024$ (24 micron pixels) with a readout noise of 4e$^-$/pixel.
With the long camera the scale at the detector was about 0.077 {\AA} /pixel.
The slit was set at $1.2\times 10$ arc seconds providing a resolving power of
$ R = \lambda / \delta\lambda$ $\approx$ 32000, as measured from the
Thorium-Argon emission lines.  The echellograms were reduced using the {\tt
MIDAS ECHELLE} package.  The background was subtracted and cosmic rays removed
with the {\tt FILTER/ECHELLE} command.  The images were not flat-fielded since
this process increases the noise in the spectra and, at the wavelengths covered
by our data, fringing is not important.  The spectra were then wavelength
calibrated using the Th-Ar lamp spectrum.  Since the observations were carried
out during bright time it was necessary to subtract the sky spectrum. For the
image covering the range 393-523 nm we had a sky spectrum taken with the same
spectrograph setting, and we subtracted that spectrum after appropriate
scaling.  For the other frame we extracted directly the sky spectrum above the
stellar spectrum, smoothed it with a gaussian 7 pixels wide to increase the
S/N, and subtracted this.  Each order was independently normalized by fitting a
spline through a set of interactively chosen points. All the orders were then
merged into a single spectrum.  These normalized spectra were shifted to rest
wavelength and over-plotted on synthetic spectra computed with the SYNTHE code
(Kurucz 1993) to perform the line identification.  This allowed us to identify
a number of probably unblended lines for which we measured equivalent widths.
The lines were measured using the {\tt IRAF} task {\tt splot } on the
non-normalized spectra, defining a local continuum and fitting a gaussian to
the line. When a line could be measured in the two different spectra or on two
adjacent orders of the same spectrum a straight average of the independent
measurements was taken.
 
Heliocentric radial velocities measured by cross-correlation  in velocity
space, using a synthetic spectrum as a template, (range 400 nm - 460 nm)  from
our two high-resolution spectra are given in Table 1.  We report there also
radial velocities measured from medium-resolution spectra obtained at the
Palomar 5.0 m and at the ESO 1.5 m telescopes.  The data is sparse, but does
not show any obvious variations of radial velocity over a long time baseline.
 
\section{Abundance analysis}
 
Our analysis relies on an extension of the ATLAS model-atmosphere grid
described in Molaro \PBal (1997) and on model atmospheres computed with the
same assumptions.  The only photometry presently available for CS 22957-027 is
broadband $UBV$.  To estimate the effective temperature from $(B-V)$ we used
two different calibrations: that of Alonso \PBal (1996) and that of McWilliam
\PBal (1995). The Alonso \PBal calibration provides also a metallicity term,
whereas the McWilliam \PBal assumes [Fe/H]$ = -3.00$. For this value of [Fe/H]
and $(B-V)=0.77$ the two calibrations coincide and yield $\rm T_{eff} = 4839
~K$.  If we adopt the BPS metallicity estimate ([Fe/H]=-3.95) the Alonso \PBal
(1996) calibration yields \PBt = 4906 K.  This star is estimated to be
unreddened by BPS -- a small reddening, of the order of 0.05 mag, would
increase the \PBt by only 100--150 K (depending on the adopted calibration).  A
concern in deriving the \PBt from $ (B-V)$ for CH and related stars is that
both bands may be affected by the strong molecular absorption.  However,
McWilliam \PBal (1995) adopt the $ (B-V)$ temperature for CS 22892-052, which
is not in contradiction with the excitation equilibrium derived.  In fact the
small line-to-line scatter found by them for Fe I is inconsistent with any
sizeable slope.  In our case, forcing the 9 unblended Fe I lines to yield a
zero slope in the abundance -- excitation energy plane would result in a \PBt ~
which is only 100 K hotter than the adopted one.  The number of lines is too
small to give much weight on this excitation temperature, therefore we decided
to adopt the photometric \PBt = 4839 K. The error on this effective temperature
may be estimated from the standard deviation of the Alonso \PBal (1996)
calibration, which is 130 K.  This error does not take into account errors on
photometry and reddening: a change of 0.01 mag in $ (B-V)$ results in a change
of 28 K in the derived \PBt .  The calibration is not very sensitive to
metallicity: a decrease of 0.3 dex in [Fe/H] implies an increase of 11 K in
\PBt, while  an increase by 0.3 dex in [Fe/H] implies a decrease of only 1 K in
\PBt.  Clearly, photometry further into the infrared would be useful to
constrain \PBt.  A microturbulent velocity of 1.7 \PBkms ~ was determined by
the requirement that the derived abundance be independent of equivalent width
for the Fe I lines.

\begin{figure}[t]
\psfig{figure=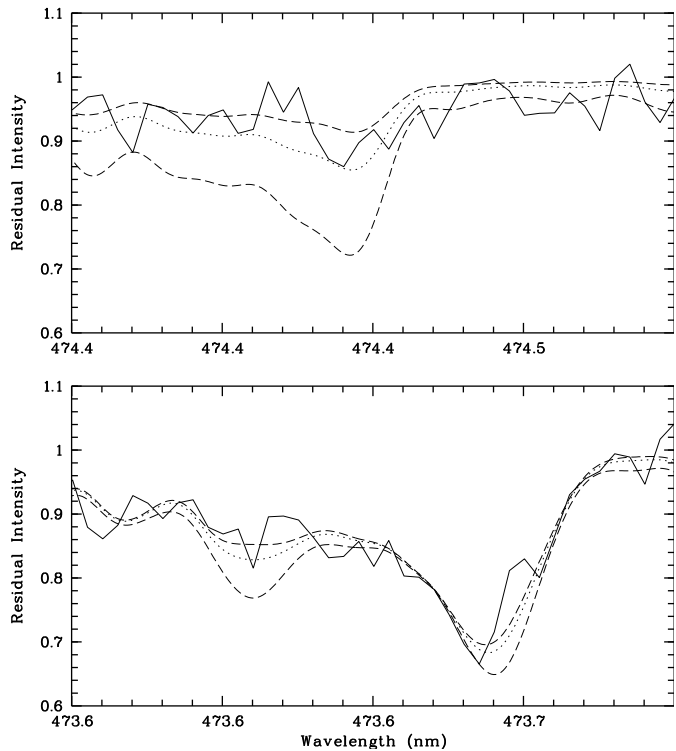,width=8.8cm,clip=t}
\caption[]{The (1,0) bands of the $\rm ^{12}C^{13}C$ (top)
and of the $\rm ^{12}C_2$  molecules. The synthetic
spectra are computed with $ \rm ^{12}C/^{13}C $ ratios
of 20,10,3 from top to bottom.}
\end{figure}

At this \PBt ~ the $ (U-B)$ color gives  little leverage on \PBg ~ , so we used
an ionization equilibrium to  estimate the surface gravity.  We could only
measure two unblended Fe II lines, but were able to measure 8 Ti I and 7
Ti II unblended lines.  We therefore used the Ti I/ Ti II equilibrium.  The Ti
I and Ti II lines yield the same abundance (within 0.02 dex) for \PBg =2.25.
The gravity is however very uncertain: a change of only 0.3 dex in the Ti II
abundance would change the gravity by one dex.  Hence we assume that the error
on  \PBg ~ is on the order of 1 dex.
 
For the unblended lines we used the WIDTH9 code to determine the abundances;
for the blended lines we used spectrum synthesis along with the SYNTHE code.
The atomic and molecular data necessary for spectrum synthesis are those of
Kurucz (1993), unless otherwise stated.  For the CH lines we placed the $gf$
values on the same scale as that used by Norris \PBal (1997a) by multiplying the
Kurucz (1993) values by 0.4.
 
\begin{table}[t]
\caption{Abundances from molecular bands}
\begin{center}
\begin{tabular}{lcccl}
\hline
  & band system & $v'$ & $v''$ & abundance\\
C$_2$   & $A^3\Pi-X^3\Pi $ & 0 & 0 & $\epsilon_C= 7.38$ ; [C/Fe]=$+2.25$\\
C$_2$   & $A^3\Pi-X^3\Pi $ & 1 & 1 & $\epsilon_C= 7.38$ ; [C/Fe]=$+2.25$\\
C$_2$   & $A^3\Pi-X^3\Pi $ & 1 & 0& $\epsilon_C= 7.33$[C/Fe]=$+2.20$\\
CH      & $A^2\Delta-X^2\Pi$ &\multicolumn2{c}{$\Delta v=-1$} & 
 $\epsilon_C= 7.13 $ ; [C/Fe]=$+2.00$\\
CN      & $ B^2\Sigma-X^2\Sigma $ & 0 & 0 & $\epsilon_N = 5.72$ ;
[N/Fe]=$+1.10$ 
\\
\hline  
\end{tabular}
\end{center}
\end{table}
 
\par
With the adopted model (4839,2.25,-3.5,1.0) and a 1.7 \PBkms ~ microturbulence,
the 17 Fe I lines, selected after excluding blends and doubtful measurements,
yield $$\rm [Fe/H]=-3.43\pm 0.12$$ where the quoted error reflects only the
line-to-line scatter.  This is somewhat higher  than the estimate of
[Fe/H]$=-3.95$ by BPS on the basis of the K index alone.  However, recent
recalibration of the CaII K / $ (B-V)$ grid by Beers \PBal (1997) obtains an
estimate of $\rm [Fe/H] = -3.59 \pm 0.20$, in close agreement with the present
value.  Thus CS 22957-027 is among the most metal--poor carbon--rich stars.
Among the similar objects which have been studied to date, only CS 29497-034 is
possibly more metal--poor than this star, depending on the true value for its
effective temperature.
 
\section{Abundance errors}
 
The abundances derived from atomic lines are summarized in Table 2 and those
from molecular lines in Table 3.
\begin{table}[t]
\caption{Abundances}
\begin{center}
\begin{tabular}{lccc}
element & $\epsilon$ & $\epsilon_\odot$ & [X/Fe] \\
\hline
C & $7.13$ & 8.56 & +2.00\\
N & $5.72$ & 8.05 & +1.10\\
Mg& $4.53$ & 7.58 & +0.38\\
Ca& $3.36$ & 6.36 & +0.43\\
Ti& $2.24$ & 4.99 & +0.68\\
Cr& $2.17$ & 5.67 & -0.07 \\
Fe& $4.05$ & 7.48 & 0.00 \\
Ni& $2.99$ & 6.25 & +0.17  \\
Sr& $-1.44$& 2.90 & -0.91 \\
Ba& $-2.23$& 2.13 & -0.93 \\
\hline
\end{tabular}
\end{center}
\end{table}

The error on the  abundances has two components: the first is related to the
presence of noise in the spectra, the second is related to the models used to
derive abundances from equivalent widths or line profiles.  The latter
includes both errors in the model-atmosphere parameters (\PBt, \PBg,
microturbulence, metallicity), and shortcomings in the model-atmosphere and
spectrum synthesis (treatment of convection, reliability of 1D-models, input
atomic and molecular data, etc.).  In this section we shall try to quantify
only abundance errors due to the noise in the spectra and to errors in the
adopted \PBt , \PBg ~ and microturbulent velocity. The results are summarized
in Table 4.
 
\subsection{Atomic lines}
\begin{table}[t]
\caption{Abundance Errors}
\begin{center}
\begin{tabular}{lcccc}
element & $\sigma_s^a$ & $\sigma_t^b$ & $\sigma_g^c$ & $\sigma_\xi^d$ \\
\hline
C &  0.20 & 0.30 & 0.4&0.1 \\
N &  0.20 & 0.10 & 0.1&0.1 \\
Mg&  0.18 &0.20  & 0.35 & 0.03\\
Ca&  0.20 & 0.15 & 0.30 & 0.05\\
Ti& 0.29 & 0.15 & 0.31 & 0.03 \\
Cr& 0.29 & 0.15 & 0.09 & 0.02  \\
Fe$^a$& 0.12 & 0.16  & 0.15 & 0.02  \\
Ni& 0.26 & 0.19 & 0.06 & 0.03  \\
Sr& 0.03 & 0.15  & 0.26 & 0.06 \\
Ba&  0.14 & 0.15 & 0.33 & 0.01 \\
\\
\hline
\end{tabular}
\end{center}
{$^a$ line to line scatter, except for C and N, see text\hfill}\\
{$^b$  change in abundance for a 130 K change in \PBt\hfill}\\
{$^c$  change in abundance for a 1 dex change in \PBg\hfill}\\
{$^d$  change in 
abundance for a 0.1 $\rm kms^{-1}$ change in microturbulent velocity\hfill}\\
\end{table}

For the blended lines we used spectrum synthesis, i.e. we computed several
synthetic spectra with different abundances of the element under scrutiny until
two ``bracketing'' synthetic spectra were found, as shown in Fig. 4. Changes of
less than 0.1 dex could not be appreciated visually, so  0.1 dex may be
considered as an order-of-magnitude estimate of the abundance error for these
lines.
 
For unblended lines one can estimate an error on the equivalent width and
compute its effect on the derived abundance.  However, for atomic species with
many lines measured we regard the line-to-line scatter more informative on the
abundance error , in spite of the fact that this is due  to the errors in the
atomic data as well as to the errors in the equivalent widths.  The error on
the equivalent width is difficult to assess, except for the isolated lines.  An
optimistic estimate of the errors on equivalent widths may be obtained from the
formula (7) of Cayrel (1988), which, for our instrumental setup, gives an error
in the range 0.6 - 0.9 pm, depending on the S/N and wavelength.  Errors in the
placement of the continuum can easily make the error on the equivalent width a
factor of two or three larger, however this is difficult to quantify. Some
improvement is expected for the lines which could measured on the two spectra
and on two adjacent orders, however this is at most a factor of two.
\par
For each element in Table 5 we report four errors:  $\sigma_s$ is the r.m.s.
error (line-to-line scatter), $\sigma_t$ is the error derived by changing \PBt
by $\pm 130$ K (the maximum of the two values was taken), $\sigma_g$ is the
error derived by changing \PBg ~ by $\pm 1$ dex and $\sigma_\xi$ derived by
changing the microturbulent velocity by $\pm 0.1$ km s$^{-1}$.  For Ca, where
formally the line to line scatter is zero, we estimate $\sigma_L$=0.2 dex at
2$\sigma$ (see above).
 
\subsection{Molecular bands}
 
For molecular bands the situation is more complex. In the first place, the
presence of the bands makes the continuum placement very difficult, in
particular  when a band completely fills one echelle order.  Different bands
and different molecules  give slightly
different abundances; the worst case is the
discrepancy between the C$_2$ Swan (0,0) band and 
the CH $\Delta v=-1$ bands, which
yield abundances differing by 0.25 dex.  However, these discrepancies could
easily be attributed to errors in the placement of the continuum or to the
adopted molecular data (a slight renormalization of the C$_2$ $gf$ values
could force the agreement).  For these reasons we believe the band-to-band
scatter in abundances to be not very informative. We performed several trials
of plausible normalizations and found that the effect on the derived abundance
was of the order of 0.2 dex; we therefore quote this value, both for C and N,
as $\sigma_s$ in Table 5.
 
There is a further source of uncertainty due to the effect of the unknown O
abundance on all the carbon-bearing molecules, and the dependence of the N
abundance derived from CN bands on the C abundance.  The O abundance is
relevant because in cool stars the formation of CO is predominant.  Since there
are no lines in our spectra suitable for the determination of an O abundance,
we assume [O/Fe]=+0.4, as observed in ``normal'' halo stars,  although in some
CH stars this ratio is lower (Vanture 1992a).  Our star is relatively hot, so
that the C abundance is not overly sensitive to the adopted O abundance; a
change in O abundance by 0.4 dex produces a change of less than 0.1 dex
in the derived C abundance, both from C$_2$ lines and CH lines.  Note, however,
that the behaviour is strongly non-linear, and much larger effects would result
by O enhancements of 1 dex or more.  The O abundance affects the CH
and C$_2$ lines in a similar way.
 
\begin{figure*}
\psfig{figure=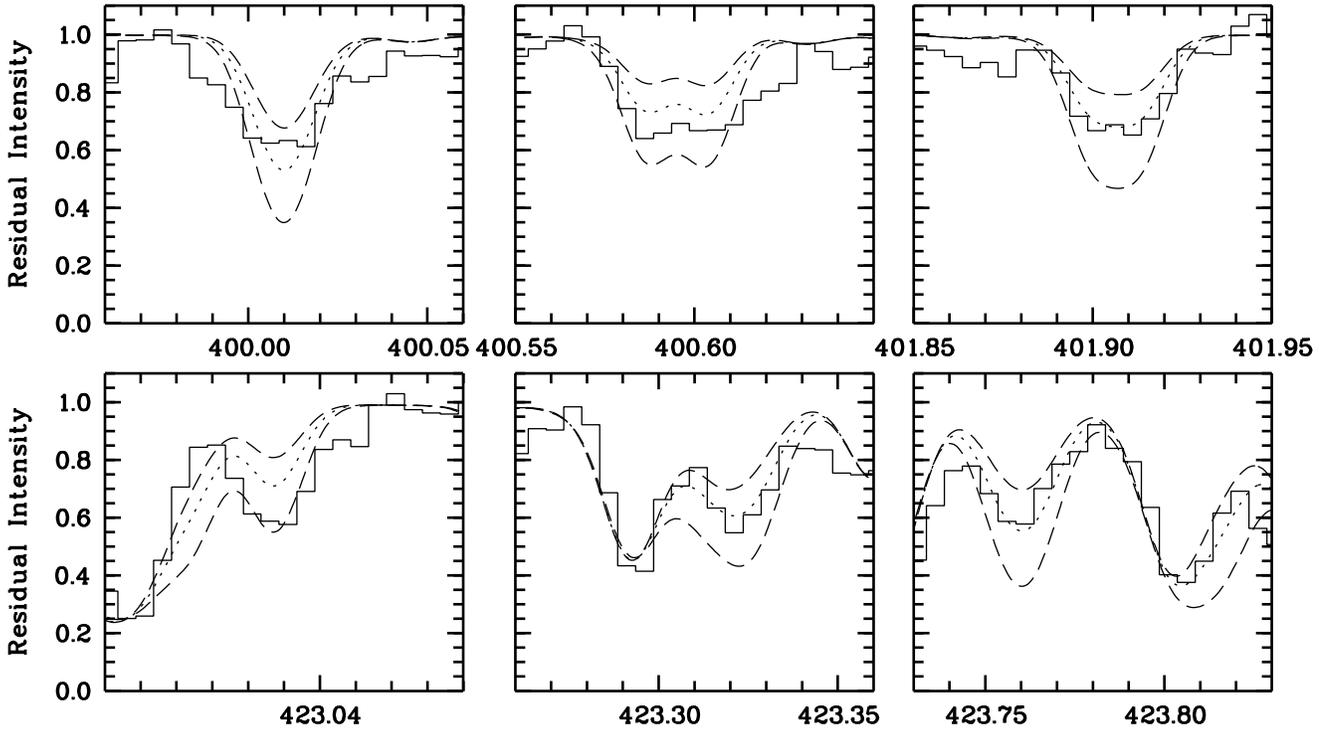,width=17.3cm}
\caption[]{Several features due to $\rm ^{13}CH$, the three
synthetic spectra superimposed as dashed or
dotted lines have been computed with $\rm ^{12}C/^{13}C$ ratios
of 20,10,3 (top to bottom), respectively.}
\end{figure*}

\section{Results}
 
The C overabundance, evident from the low-resolution spectroscopy, is
confirmed.  In Fig. 1 the C$_2$ (0,0) Swan bandhead, together with the Mg I b
triplet, is shown in order to illustrate the significant carbon enhancement in
CS 22957-027.  The superimposed synthetic spectrum has been computed assuming
[C/Fe]=+2.25 and [Mg/Fe]=+0.38. Our results are summarized in Table 4.  There
we report the C abundance derived from CH lines, since these affect the
observed spectrum more than C$_2$ lines.  We have also been able to detect the
(1,0) Swan band of the $\rm ^{12}C^{13}C$ molecule, shown in Fig. 2, together
with the same bandhead for $\rm ^{12}C_2$ (lower panel).  For the spectrum
synthesis of the $\rm ^{12}C^{13}C$ band we employed the same molecular data as
Aoki \& Tsuji (1997), namely: wavelengths from Pesic \PBal (1983), band
oscillator strength of $1.27\times 10^{-2}$, derived from the lifetime
measurement of Naulin \PBal (1988), and H\"onl-Condon factors from Danylewych
\PBal (1974).  The $\rm ^{12}C/^{13}C$ ratio derived from this band is $\approx
10$, thus requiring a considerable enrichment in $\rm ^{13}C$.
 
Norris \PBal (1997b) noted that $\rm ^{13}CH$ lines are present in the spectrum
of CS 22957-027. Some of the $\rm ^{13}CH$  identified in our spectrum are
shown in Fig. 3.  The figure also shows superimposed synthetic spectra and it
may be appreciated that the isotopic ratio lies between 3 (equilibrium value
for the CN cycle) and 20, with a preferred rather low $\rm ^{12}C/^{13}C$ value
around 10, in concert with the results from the Swan (1,0) band.  In the
spectral synthesis the wavelengths of some of the $\rm CH$ lines,  listed as
``predicted'' in Kurucz (1993), have been changed slightly to bring
observations and computations into better agreement. The modified wavelengths
are listed in detail in Table 6 .
\par
Our spectra of CS 22957-027 show a strong absorption feature at 401.905 nm, 
which
has a good wavelength coincidence with the Th II 401.9129 nm resonance line
(Fig. 3) which was used by Cowan \PBal (1997) in CS 22892-052 for
nucleo-chronology.  However, in CS 22957-027 two $\rm ^{13}CH$ lines fall in
this spectral region, and we believe they are the main contributors to the
feature, but some adjustment of the predicted wavelengths is necessary (see
Table 6).  If the whole feature is interpreted as due to Th II only, it implies
a very high Th abundance ([Th/H]=+2.3, using  log gf=$-0.27$ of Lawler \PBal
~1990), but spectrum synthesis with this high Th abundance is not satisfactory
and fails to reproduce the blue wing of the feature.  The FWHM of the feature
is 0.032 nm, while typical widths of unblended Fe I lines in our spectra are
0.02 nm, revealing that the feature is a blend rather than a single line.  The
lack of detection of Eu lines suggests that the feature at 401.9 nm is not
mainly due to Th, since  the $r-$process requires both elements be synthesized
at the same time.
 
The presence of $\rm ^{13}CH$ lines blending the Th II line is of relevance for
the use of this Th II line in these stars for nucleo-chronology when the $\rm
^{13}C$ carbon isotope is enhanced at the levels observed in this star.
 
\begin{figure}
\psfig{figure=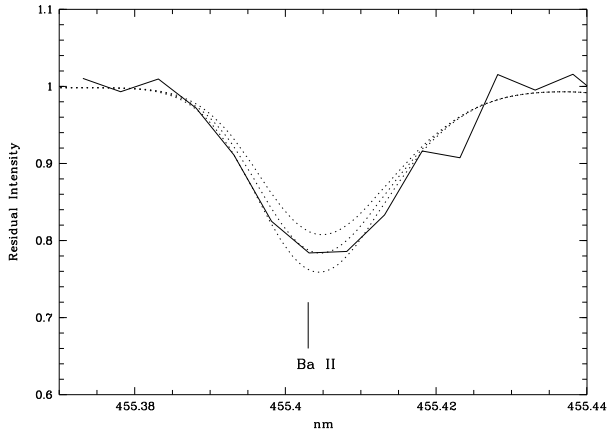,width=8.8cm,angle=-90}
\caption[]{The Ba II resonance line. The dotted lines  synthetic spectra
with $\epsilon_{\rm Ba}= -2.00 $  
$\epsilon_{\rm Ba}= -2.10 $, and $\epsilon_{\rm Ba} = -2.20 $ }
\end{figure}
 
The N abundance is deduced from the CN $B^2\Sigma-X^2\Sigma$ violet system and
in particular from the (0,0) bandhead at 388.3 nm.  It depends on the adopted C
abundance: for [C/Fe]=+2.00, as indicated by the CH lines, the N
enhancement is  [N/Fe]=+1.1, which implies C/N $\approx 26 $.  
In
``classical'' CH stars the C/N ratio is always greater than 1, in the range
2--6, as is expected from products of partial CN cycling (Vanture 1992a).  
N is found to be
enhanced over iron in all the other very-metal-poor C-enhanced stars (Norris
\PBal 1997a ; Barbuy \PBal 1997) with C/N ratios in the range 4--15.
The [C/N] ration in CS 22957-027 are compared with that of
other metal-poor carbon-rich stars in Fig.  5.
 
\begin{table}[t]
\caption{Modified wavelengths for some CH lines}
\begin{center}
\begin{tabular}{lcl}
$\lambda$ & label & $\lambda$ \\
Kurucz93    &       & adopted\\
\hline
399.9661& 106.00X00F2     B00E2   13 & 400.0061\\
399.9738& 106.00X00E1     B00F1   13 & 400.0138\\
400.0996& 106.00X00F2     B00E2   12 & 400.1266\\
400.5666& 106.00X00E2     B00E2   13 & 400.5866\\
400.5840& 106.00X00F1     B00F1   13 & 400.6040\\
401.6749& 106.00X00F2     B00E2   12 & 401.7149\\
401.8765& 106.00X00E2     B00E2   13 & 401.9000\\
401.8936& 106.00X00F1     B00F1   13 & 401.9136\\
423.1222& 106.00X00F2     A00F2   13 & 423.1422\\
423.1281& 106.00X00E1     A00E1   13 & 423.1481\\
423.1294& 106.00X00E1     A00F2   13 & 423.1494\\
423.3039& 106.00X01E2     A01E2   13 & 423.3239\\
423.3051& 106.00X01F2     A01F2   13 & 423.3251\\
423.3088& 106.00X01E1     A01E1   13 & 423.3288\\
423.0170& 106.00X00F1     A00F1   13 & 423.0270\\
423.0183& 106.00X00F1     A00E2   13 & 423.0283\\
423.7354& 106.00X00F2     A00F2   13 & 423.7554\\
423.7445& 106.00X00E1     A00E1   13 & 423.7645\\
423.7450& 106.00X00E1     A00F2   13 & 423.7650\\
437.0416& 106.00X00F2     A00F2   13 & 437.0636\\
437.0421& 106.00X00F2     A00E1   13 & 437.0651\\
\hline
\end{tabular}
\end{center}
\end{table}

The violet CN system  poses some problems with abundance determinations, for
example Norris \& Da Costa (1995) found that its use yields abundances which
are 0.5 dex lower than the blue CN band around 420 nm ($B^2\Sigma-X^2\sigma
~~\Delta v = -1$). We do not have spectra for a metal-poor star with
well-determined N abundance, e.g. from NH lines, to allow us a renormalization
of the $gf$ values of the kind used by Norris \PBal (1997a). However we expect
that such a procedure would increase our derived N abundance by about 0.5 dex.
 
The data on Mg and Ca shows an [$\alpha$/Fe] ratio larger than solar, as
observed in halo stars of this low metallicity, not showing evidence of any
peculiarities.  Furthermore, the Ti abundance supports these findings, although
it is slightly higher than ``normal.'' Note, however, that the scatter of Ti
abundances is on the order of 0.3 dex.  Ni and Cr closely track Fe,  as is
expected for  elements produced under conditions of nuclear statistical
equilibrium.
 
Sr and Ba in CS 22957-027 are deficient with respect to iron ([Sr/Fe] $=-0.91$,
[Ba/Fe]$ =-0.93$) -- the Ba II resonance line at 455.4 nm is shown in Fig. 4 as
an example.  The Nd II and Eu II resonance lines at 411.0472 nm and  412.97 nm,
respectively, are not detected.  A deficiency appears to be present in all
$n-$capture elements, in contrast to the large overabundance observed in CH and
Ba stars.  Figure 5 shows that CS 22957-027 has [Ba/C] about 2.5 dex lower than
the other five very-metal-poor CH stars (CS 22892-052 McWilliam \PBal 1995,
Sneden \PBal 1996; CS 22948-027, CS 29497-034 Barbuy \PBal 1997; LP 706-7 and
LP 625-44 Norris\PBal  1997a).
 
\begin{figure}[t]
\psfig{figure=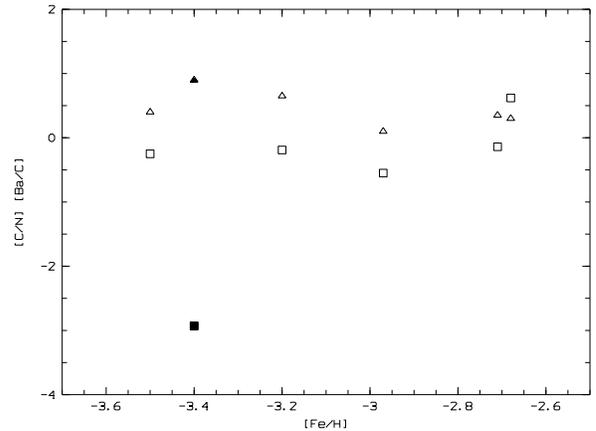,width=8.8cm,angle=-90}
\caption[]{ 
[C/N] (triangles) and [Ba/C] (squares) for the very metal poor CH stars.
Filled symbols refer to CS22957-027. 
}
\end{figure}

\subsection{Comparison with Norris \PBal (1997b)}

After submitting this work  we became aware of
the analysis of Norris \PBal (1997b) of this same star.
The atmospheric parameters derived by Norris \PBal (1997b)
are very close to those adopted by us, \PBt = 4850, \PBg =1.9
and microturbulent velocity of 1.5 \PBkms.
The agreement of \PBt is not surprising, since both 
are derived from the $(B-V)$ colour, albeit with 
different calibrations, but
the agreement of surface gravities is, on the other hand, quite significant.
Norris \PBal (1997b) derived \PBg ~from the FeI/FeII ionization
equilibrium, while we derived it from the TiI/TiII equilibrium,
thus the two results strengthen each other and the quoted
1 dex error for our \PBg ~is probably an overestimate.

The metallicity of CS22957-027 derived by
Norris \PBal (1997b) is [Fe/H]$=-3.38\pm 0.13$, which is
in remarkably good agreement with our result. 
Norris \PBal (1997) use model-atmospheres based on Bell \PBal (1976), 
whereas we use
ATLAS model-atmospheres with no convective overshooting and
$\alpha-$ element enhanced ODFs. 
The abundances of the other elements 
in common are also in substantial agreement
and so is the $\rm ^{12}C/^{13}C$.
Exceptions are N, Ti and Sr for which the difference exceeds
0.3 dex. We have already commented 
on the problems with  N determinations, 
we note here that the suggested increase of 0.5 dex in our
N abundance would bring our result closer to that of Norris \PBal (1997b).
For Ti the difference may be attributed to the different set of
lines
used, in fact we have only 4 Ti II lines in common. 
The discrepancy for Sr can be explained by the fact that
Norris \PBal (1997b) determine the abundance from the equivalent width.
Our synthetic spectrum reveals that many
weak CN lines lower the continuum. The total
equivalent width  of the feature 
on our spectra is 7.7 pm, in agreement with the value of 8.0 pm
of Norris \PBal (1997b). 

The radial velocity measured by Norris \PBal (1997b) is of 
$71.3\pm 0.2$ \PBkms ~(internal error) and again does not
support any evidence of variability. Their spectra were
taken on  August 5$^{th}$ 1996 and provide a short 
time baseline if compared with ours.

\section{Discussion}
 
We have provided evidence that CS 22957-027, with [Fe/H]$=-3.43$, is one of the
most metal-deficient carbon stars.  The  $\rm ^{12}C/^{13}C$ is $\approx$ 10,
indicating an enhancement of $\rm ^{13}C$, and the {\it s}-process elements are
at the values found in non carbon stars with similar metallicities.  Of the
numerous classes of carbon stars, at least three bear some resemblance to CS
22957-027: CH stars, early R stars, and AGB stars (late R and N stars).  In the
following we discuss the possible membership of CS 22957-027 to these groups,
but conclude that this star is unique, and does not match entirely the
properties of any of these groups of stars.
 
\par
CH stars are metal-deficient stars characterized by strong G bands, with
enhancements of $s-$process elements at the level observed in the AGB stars.
Typical [$\it s$/Fe] in the CH stars are from 0.6 to 1.3 dex.  Since their
inferred luminosities are too low for AGB stars, a model of mass transfer from
a carbon star, presently evolved to white dwarf, to  its companion, the CH
star, has been proposed.  In fact the majority of the CH stars have been shown
to be binaries by McClure \& Woodsworth (1990).  Most of the CH stars have $\rm
^{12}C/^{13}C$ ratios around 10, but a small fraction exhibit very high values,
in excess of 500.  The low  $\rm ^{12}C/^{13}C$ values are a factor of 3 higher
than the metal-poor RGB, which exhibit a peak at the CN-cycle equilibrium value
($\rm ^{12}C/^{13}C=$  3-4). The high values are much higher than that
observed in the population I carbon stars of N-type (Aoki and Tsuji 1997).  The
two different isotopic ratios probably reflect different astrophysical origins.
Those with very high $\rm ^{12}C/^{13}C$ can be explained, like the carbon
stars of N-type, by transfer of material which contains $\rm ^{12}C$  dredged
up from the He-burning layer.  The CH stars with low $\rm ^{12}C/^{13}C$
require a transfer of $\rm ^{12}C$, but also  a dredge-up of the products of
internal CNO-cycle in the accreting star to lower the $\rm ^{12}C/^{13}C$.
Another possibility is that mass transfer involves material where  the $\rm
^{12}C/^{13}C$ has already been lowered by hot-bottom burning  (Renzini \&
Voli 1981 ).  This latter binary-star scenario can explain the carbon and
nitrogen enhancement and the low $\rm ^{12}C/^{13}C$ ratio 
found in CS22957-027,
however the  modest amount of $s-$process elements found by us is rather
peculiar among CH stars.  On the other hand this deficiency is not unusual for
very-low-metallicity stars ([Fe/H]$\le -2.5$ ) which are {\em not} C- or N-
enhanced.  Inspection of Fig. 7a of Ryan \PBal (1996), where the $s-$ process
element abundances versus metallicity for very metal-poor stars are plotted,
shows that CS22957-027 falls in the region where the heaviest concentration of
points lie.  In addition to the normal abundances of neutron-capture elements,
the ratio [Ba/Sr] is solar, at variance with what observed in CH stars, where
the heavy $s-$process (such as Ba)
to light $s-$process (such as Sr)
elements ratio ([hs/ls]) is
super-solar (ranging from about 0.5 to 2 dex, with an average value of 0.9;
Vanture, 1992b).  It has been also suggested by Aoki and Tsuji (1997) that the
CH stars may be the population II counterparts of Ba stars. If this is so, then
CS22957-027 should not belong to the CH group.
\par
Can it be that CS22957-027 is on the thermally pulsating AGB (TP-AGB) and thus
self-enriched by the third dredge-up?  Our derived surface gravity places the
star below the TP-AGB luminosities, and it is therefore unlikely that it has
undergone self-enrichment.  Evolutionary arguments also
do not support the AGB hypothesis.  Very-low-mass stars do not undergo the
third dredge-up, although the exact value of the minimum main-sequence mass for
the third dredge-up to occur is not known precisely; Marigo \PBal (1996) find
that the minimum mass for their metal-poor models (Z=0.008) to undergo
dredge-up is 1.1$\rm M_\odot$, while Gustafsson \& Ryde (1996) suggest  $\rm
1.2 M_\odot$.  The evolutionary time--scales for a metal-poor star of mass
between 1 and 1.2 M$_\odot$ to reach the TP-AGB are on the order of 7--4 Gyr
(Boothroyd \& Sackman 1988), which is almost a factor of two less than for a
solar-metallicity star.  If the very-metal-poor stars were formed some 10 Gyr
ago, we do not expect any of them to be presently on the TP-AGB after
experiencing dredge-up.  Less-massive stars could be on the TP-AGB, but should
not have experienced the third dredge-up.  A relatively young (3--10 Gyr)
metal-poor population has been identified  by Preston \PBal (1994), the
so-called Blue Metal Poor (BMP) stars, and these stars could  be presently
observed on the AGB.  Thus it might be possible to understand the chemical
peculiarities we observe as due to self-enrichment only if CS 22957-027 is more
luminous than we derived and is relatively young, as it could be if it belongs
to the cool end of the BMP population. In this case the lack of
enhancement of $s-$ process elements poses some difficulties.
\par
Another class of stars which bears some resemblance to CS 22957-027 are the
early-R stars. These are  a class of carbon  stars without enhancement of {\it
s}-process elements (Dominy 1984).  The solar abundance for {\it s}-elements in
the early-R stars is in sharp contrast with other related stars (N, S, Ba
giants, Ba dwarfs, CH, sgCH, dwarf carbon stars) which all show significantly-
enhanced abundances of the $s-$process elements relative to iron (Green and
Margon 1994).  In early-R stars the $\rm ^{12}C/^{13}C$ ranges from near 4 to
15.  A sample of early-R stars has not shown evidence of binarity after a 16
year-baseline monitoring of radial velocity (McClure 1997a), ruling out the
binarity, in contrast to what has been found for the  CH stars.  Typical
metallicities for the early-R stars are solar or slightly metal-deficient, in
the range -$\rm 0.6 < [Fe/H] < 0.0$.  
The  distinguishing features of early-R stars
from the CH stars are the non-enhancement of $s-$process elements and the
non-binarity of the early-R stars.  The fact that no metal-poor early-R stars
are known may not imply that such stars do not exist. In fact, they may have
escaped detection and misclassified as CH stars, the only safe discriminating
criterion being the abundance of the $s-$ elements.  There is no agreement as
to the origin of early-R stars, but the most viable hypothesis is that of
mixing of C produced by explosive helium-core flash  at the tip of the RGB
(Dominy 1984).  More exotic hypothesis such as the coalescence of a binary
system have been also suggested (McClure 1997a).
 
CS22957-027 appears to be similar to the early-R stars, except for the much
lower metallicity.  Early-R stars show [C/H] in the range +0.1 to +0.8, or
[C/Fe] in the range +0.2 to +1.0, while CS22957-027 has [C/H]=-1.4 and
[C/Fe]=+2.0,  thus showing a lower carbon content, albeit with a higher [C/Fe].
The early-R stars  require a  conspicuous carbon enrichment, at least on the
order of 30\% of the solar carbon.  If the mechanism responsible for the
formation of early-R stars is the helium-core flash one could expect similar
amounts of carbon to be mixed, whatever the metallicity. This would rule out
the possibility of an early-R star with a low carbon content.  However it is
not inconceivable that the amount of carbon mixed during the He-flash depends
also on the initial metallicity, being lower for lower-metallicity stars.
 
Thus, CS 22957-027 appears unique among the carbon stars at very low
metallicity, requiring the [C/Fe] and [N/Fe] enhancements to be  created by a
process which does not lead also to an $s-$process element enrichment.  The
few other carbon stars at low metallicity presently known have thus-far proven
to be enhanced in $s$--process elements, like the  classical CH and Ba stars.
They may have a similar nature to the latter, and mass transfer in a binary
system may be able to explain their origin.  However, the $r-$process element
enrichment of CS 22892-052 needs another explanation. None of the scenarios
so far proposed appears entirely satisfactory.  Ryan \PBal (1996) already
pointed out that the very-metal-poor carbon stars probably encompass objects
with more than one origin.  CS 22957-027 is another object which poses some
problems to common wisdom on carbon--rich objects.  Its properties   could be
described by saying that it is a Pop. II R0-R5 star or a peculiar CH star,
neither of which has been previously observed.  Further scrutiny of other
carbon-enhanced stars is clearly warranted and can provide important clues
towards an understanding of nucleosynthesis processes in the early Galaxy.
 
\begin{acknowledgements}
This research was partially supported by Collaborative NATO grant No. 950875.
TCB acknowledges support from NSF 
grant AST 95-29454 awarded to Michigan State
University. We wish to thank B. Barbuy for helpful discussions on the synthesis
of molecular bands and G. Bono for helpful discussions on AGB stars.
\end{acknowledgements}


\begin{thebibliography}{1}
\bibitem{ }\PBa  Alonso A., Arribas S., Mart\PBi nez-Roger C.:1996 313 873 
\bibitem{ }\PBa Aoki W., Tsuji T.:1997 317 845
\bibitem{ }\PBa Barbuy, B., Cayrel, R., Spite, M., Beers, T.C.,
Spite, F., Nordstr\"om, B.,  Nissen P.E.:1997 317 L63
\bibitem{ }\PBaj Beers T.C., Preston G.W.,  Shectman S.A.:1985 90 2089
\bibitem{ }\PBaj Beers T.C., Preston G.W.,  Shectman S.A.:1992 103 1987
\bibitem{ } Beers T.C., Rossi, S., Norris, J.E., Ryan, S.G., 
Shefler, T., 1997, in preparation 
\bibitem{ }\PBasupl Bell R.A., Eriksson K., Gustafsson B., Nordlund A.:1976
23 37
\bibitem{ }\PBapj Boothroyd A.I.,  Sackman I.J.:1988 328 671
\bibitem{ }  Cayrel R., 1988 in
Cayrel de Strobel G., Spite M., eds, Proc. IAU Symp. 132, 
The Impact of Very High S/N Spectroscopy
on Stellar Physics. Kluwer,
Dordrecht, p. 345
\bibitem{ }\PBapj Cowan J.J., McWilliam A., Sneden C., 
 Burris D.L.:1997 480 246
\bibitem{ }\PBprsoc Danylewych L.L:, Nicholls R.W.:1974 399 197
\bibitem{ }\PBapjsupl Dominy J. F.:1984 55 27
\bibitem{ }\PBa Goriely S., Arnould M.:1997 322 L29
\bibitem{ }\PBapj Green P.J., Margon B.:1994 423 723
\bibitem{ } Gustafsson B., Ryde N., 1996  IAU Symposium 177
`` The Carbon Star Phenomenon '' {astro-ph/9610261}
\bibitem{ } Kurucz R.L. 1993, CD-ROM 13, 18
\bibitem{ }\PBnat Lawler J.E., Whaling W.,  Grevesse N.:1990 346 635
\bibitem{ }\PBapj Luck R.E.,  Bond H.E.:1982 259 792
\bibitem{ }\PBpasp McClure R.D.:1984 96 117 
\bibitem{ }\PBpasp McClure R.D.:1997a 109 256
\bibitem{ }\PBpasp McClure R.D.:1997b 109 536 
\bibitem{ }\PBapj McClure R.D.,  Woodsworth A.W.:1990 352 709 
\bibitem{ }\PBaj McWilliam, A., Preston, G.W., Sneden, C.,
 Searle L.:1995 109 2757
\bibitem{ }\PBa Marigo P., Bressan A., Chiosi C.:1996 313 545
\bibitem{ }\PBa Molaro P., Bonifacio P., Castelli F.,  
Pasquini L.:1997 319 593
\bibitem{ }\PBchphyl Naulin C., Costes M., Dorthe G.:1988 143 496
\bibitem{ }\PBapj Norris J.E., Da Costa G.S.:1995 447 680  
\bibitem{ }\PBapj Norris J.E., Ryan S.G.,  
   Beers T.C.:1997a 488 350 
\bibitem{ }\PBapj Norris J.E., Ryan S.G.,  
   Beers T.C.:1997b 489 L169 
\bibitem{ }\PBjmols Pesic D.S., Vujisoc B.R., Rakotoarijimy D.,
Weniger S.:1983 100 245 
\bibitem{ }\PBaj Preston G.W., Beers T.C.,  Shectman S.A.:1994 108 538
\bibitem{ }\PBa Renzini A., Voli M.:1981 94 175
\bibitem{ }\PBapj  Ryan S.G., Norris J.E.,  Beers T.C.:1996 471 254 
\bibitem{ }\PBapj Sneden, C., McWilliam, A., Preston, G.W.,
Cowan, J.J., Burris, D.L.,  Armosky, B.J.:1996 467 819
\bibitem{ }\PBaj Vanture, A.D.:1992a 104 1986
\bibitem{ }\PBaj Vanture, A.D.:1992b 104 1997
\bibitem{ }\PBapjsupl Woosley S.E., Weaver T.E.:1995 101 181
\end{thebibliography}
\end{document}